\documentclass[a4paper,12pt]{article}
\usepackage[top=3cm]{geometry}
\usepackage{latexsym}
\usepackage{amsfonts}
\usepackage{amsmath}

\usepackage{color}
\usepackage{amsthm}
\usepackage{braket}
\usepackage{tabularx}
\usepackage{makeidx}
\usepackage{mathtools}
\usepackage{enumerate}
\usepackage{framed}
\usepackage{bbold}
\usepackage{tikz-cd}

\newcommand{\C}{\mathbb{C}}
\newcommand{\CP}{\mathbb{CP}}

\newcommand{\T}{\mathbb{T}}
\newcommand{\Z}{\mathbb{Z}}
\newcommand{\p}{\partial}

\newcommand{\CA}{{\mathcal A}}

\newcommand{\Lie}{{\mathcal L}}
\newcommand{\CO}{{\mathcal O}}

\newcommand{\tr}{\mathrm{tr}}

\newcommand{\be}[1]{\begin{equation}\label{#1}}
\newcommand{\ee}{\end{equation}}
\newcommand{\bea}[1]{\begin{eqnarray}\label{#1}}
\newcommand{\eea}{\end{eqnarray}}
\newcommand{\bean}[1]{\begin{eqnarray*}}
\newcommand{\eean}{\end{eqnarray*}}
\newtheorem{thm}{Theorem}
\newtheorem{defn}{Definition}
\newtheorem{propn}{Proposition}
\newtheorem{lemma}{Lemma}
\newtheorem{coroll}{Corollary}

\newcommand{\parz}{\partial_z}

\newcommand{\paru}{\partial_u}

\newcommand{\parrh}{\partial_{\rho}}

\title{Meromorphic Painlev\'e III transcendents and the Joukowski correspondence}
\author{Andrea Ferrari\footnote{\texttt{andrea.ferrari@maths.ox.ac.uk}} \;  
\&  Lionel Mason\footnote{\texttt{lmason@maths.ox.ac.uk}}
\\
\small{        The Mathematical Institute,} \\\ \small{AWB, ROQ, Woodstock Rd, Oxford OX2 6GG.
    }   }
\date{}
\begin{document}
\maketitle
\begin{abstract}
We study a twistor correspondence based on the Joukowski map  reduced from one for stationary-axisymmetric self-dual Yang-Mills and adapt it to the Painlev\'e III equation.  A natural condition on the geometry (axis-simplicity) leads to solutions that are meromorphic at the fixed singularity at the origin. We show that it also implies a quantisation condition for the parameter in the equation.  From the point of view of generalized monodromy data, the condition  is equivalent to triviality of the Stokes matrices and half-integral exponents of formal monodromy. We obtain canonically defined representations in terms of a Birkhoff factorization whose entries are related to the data at the origin and the Painlev\'e constants.
\end{abstract}

\section{Introduction}
The self-dual Yang-Mills equations provide a paradigm of complete integrability by virtue of their twistor correspondence \cite{Ward:1977ta}.  This expresses local solutions in terms of essentially free holomorphic data on an auxiliary complex manifold, twistor space. 
Symmetry reductions lead to many of the most basic integrable systems and their integrability  can be understood  via the reduction of this twistor correspondence \cite{Mason:1996}.    
When the self-dual Yang-Mills equations are stationary and axisymmetric, the reduced twistor correspondence is based on a parametrised family of  Joukowski (or Zhukovski) transformations \cite{Mason:1988}. 
 When the gauge group is SL(2), interesting reductions of this type include the Ernst equations of general relativity, and when a further symmetry is imposed, the third and sixth Painlev\'e equations. 
 
This paper started as an exploration of the connections between this Joukowski correspondence and the Quantum Spectral Curve of Gromov, Kazakov and Volin \cite{Gromov:2014caa}.  This is based on the same geometry and produces quantum field theoretic anomalous dimensions as functions of just one variable, the coupling constant, via quantum rather than classical integrability.  Such objects are naturally expected to be holomorphic near zero coupling.  In \cite{Mason:1996}, based on  \cite{Ward:1982bf}, solutions of the stationary axisymmetric self-dual Yang-Mills equations were studied that had well-defined meromorphic behaviour at the axis, termed \emph{axis-simple}.  The radial coordinate plays the role of the coupling constant in the quantum spectral curve, and so anomalous dimension should relate to this axis-simple class given that they should be in fact holomorphic near the origin. There are additional ingredients in the computation of anomalous dimensions, not straightforwardly reflected in these constructions. The Painlev\'e III equation, however, appears to be a close analogue\footnote{Especially to the BPS limits considered in \cite{Gromov:2014bva,Gromov:2015dfa}.}.

This was our motivation for the exploration given here for the Painlev\'e III equation in this axis-simple case. These results seem to be of interest in their own right and so this paper provides a separate study of these meromorphic solutions to the Painlev\'e III equation. We find a quantization of parameters that arises from the axis-simple condition and see a simplified Riemann-Hilbert problem for constructing such solutions. From the point of view of isomonodromic deformations \cite{Jimbo:1981zz,Its:1986,Its:2011}, we show that the axis-simple condition is equivalent to the triviality of the Stokes matrices together with half-integral exponents of formal monodomy.

This paper is structured as follows. In the remainder of this introduction, we introduce the $P_{III}$ equation and state our main results. In Section~\ref{sec:JukowskiCorrespondence} we review the reduced twistor correspondence for stationary axisymmetric self-dual Yang-Mills equations which gives solutions with meromorphic behaviour at the origin. 

In Section~\ref{sec:Proof} we adapt the construction to Painlev\'e III and show that this leads to a quantisation of the parameters and prove our main theorem. We go on to characterize the axis-simple condition in terms of generalized monodromy data. In Section~\ref{sec:WardAnsatz} we reconstruct some explicit solutions that reproduce the classical transcendental solutions of Painlev\'e III meromorphic at $\rho = 0$.

\subsection{The Painlev\'e III equation}

The Painlev\'e equations are second order ordinary differential equations whose only movable singularities,  (\emph{i.e.}\ the singularities of their solutions whose locations depend on the initial conditions) are poles. The equations have received much attention in mathematical physics over the years (ref.~\cite{Ince:1956,McCoy:1976cd, Ablowitz:1991,Zamolodchikov:1994uw,Tod:1995,Alday:2009yn,Guest:2015kua}). 
 They can all be viewed as symmetry reductions of the  $\mathrm{SL}(2, \mathbb{C})$ SDYM equations  by certain three dimensional abelian subgroups of the conformal group \cite{Mason:1993, Mason:1996}. Painlev\'e III,  $P_{III}$, can in particular be obtained from the stationary axisymmetric equations  by imposing an additional translational symmetry along the axis \cite{Calvert:1996}.

The third Painlev\'e equation is a family of equations parametrized by four complex parameters $(\alpha , \beta , \gamma , \delta )$:
\begin{equation}
f^{''} =  \frac{(f^{'})^2}{f} - \frac{f^{'}}{\rho} + \frac{1}{\rho}(\alpha f^2 + \beta) + \gamma f^3 + \frac{\delta}{f} \, .
\label{PainleveIII}
\end{equation}
For all  $(\alpha , \beta , \gamma , \delta )$, the equation is a meromorphic ODE with a simple pole at its fixed singularity at $\rho=0$.

It is customary (see \cite{Okamoto:1987}, \cite{Guest:2015kua}) to distinguish four different classes:
\begin{equation}
\begin{split}
P_{III}(D_6)&= \{ (\alpha, \beta, \gamma, \delta) \ | \ \gamma \delta \neq 0\} \, , \\
P_{III}(D_7)&= \{ (\alpha, \beta, \gamma, \delta) \ | \ \gamma = 0,\  \alpha  \delta \neq 0, \ \mathrm{or} \ \delta = 0, \ \beta \gamma \neq 0 \} \, \\
P_{III}(D_8)&= \{ (\alpha, \beta, \gamma, \delta) \ | \ \gamma = \delta = 0 , \ \alpha \beta \neq 0\} \, , \\
P_{III}(Q)&= \{ (\alpha, \beta, \gamma, \delta) \ | \ \alpha = \gamma = 0 \ \mathrm{or} \ \beta =  \delta = 0\} \, .
\end{split}
\end{equation} 
Other commonly used parameters are
\begin{equation}
\alpha = -8n, \quad \beta = 8(m-k), \quad \gamma = 16 l^2, \quad \delta = -16 k^2,
\label{parameterchange}
\end{equation} 
in terms of which the above families become
\begin{equation}
\begin{split}
P_{III}(D_6)&= \{ (l, k, m, n) \ | \ kl \neq 0\} \, , \\
P_{III}(D_7)&= \{ (l, k, m, n) \ | \ l=0,\  kn \neq 0, \ \mathrm{or} \ k=0, \ lm \neq 0 \} \, , \\
P_{III}(D_8)&= \{ (l, k, m, n) \ | \ l = k = 0 , \ n \neq 0, \ m \neq k \} \, , \\
P_{III}(Q)&= \{ (l, k, m, n) \ | \ n=l = 0 \ \mathrm{or} \ m =  k = 0\} \, .
\end{split}
\end{equation} 
Rescaling  $f$ and $\rho$ reduces the number of essential free parameters in each class to $2, \ 1, \ 0, \ 1$ respectively. The most familiar case is the scaling reduction of the Sinh-Gordon equation which can be expressed as either $D_6$ with $\alpha=\beta=0$ or  $D_8$.

\subsection{Monodromy preserving deformations} \label{MonodromyData} 
 It is well-known 
that the Painlev\'e III equation describes the isomonodromic deformations of a $2 \times 2$ linear system of equations
\begin{equation}\label{linearsystem}
\frac{\mathrm{d} Y (\lambda)}{\mathrm{d}\lambda} = \left( \frac{A_{0,1}}{\lambda^2} + \frac{A_{0,0}}{\lambda} - A_{\infty , 0}  - A_{\infty ,1} \lambda \right)  Y (\lambda) \,
\end{equation} 
with Poincar\'e rank 1 at the irregular singularities $\lambda=0$ and $\lambda = \infty$ \cite{Flaschka:1980,Jimbo:1981zz, Its:2011}. We briefly review the generalized monodromy data and establish some notation. For simplicity, in what follows we only consider the case where all matrices $A_{i,j}$ appearing in the above equation are diagonalisable.  Near the singularities the system has formal solutions
\begin{equation}\label{formalsolutions}
Y(\lambda)^{(0,\infty)} = G^{(0,\infty)} \hat Y (\lambda)^{(0,\infty)} e^{T^{(0,\infty)} (\lambda)} \, .
\end{equation}
Here
\begin{equation}
\hat Y(\lambda)^{(0)} = 1 + y_1^{(0)} \lambda + y_2^{(0)} \lambda^2 + \ldots
\end{equation}
and
\begin{equation}
\hat Y(\lambda)^{(\infty)} = 1 + y_1^{(\infty)} \lambda^{-1} + y_2^{(\infty)} \lambda^{-2} + \ldots
\end{equation}
are power series around the irregular singularities that converge in wedges centered on them, the so-called \emph{Stokes sectors} $\mathcal{S}_j^{(0,\infty)}$  whose details will not be needed  here. The $G^{(0,\infty)}$s are constant matrices that diagonalise $A_{0,1}$ and $A_{\infty,1}$, and
\begin{align}
T^{(0)} &= -T^{(0)}_{-1} \lambda^{-1} + T^{(0)}_{0} \log (\lambda)  \\
T^{(\infty )} &= -T^{(\infty )}_{-1} \lambda + T^{(\infty)}_0 \log \left(\frac{1}{\lambda}\right) \, ,
\end{align}
where the $T^{(0,\infty)}_{-1,0}$s are diagonal. For each sector $\mathcal{S}_j^{(0,\infty)}$ labelled by $j$, there are convergent solutions $Y_j^{(0,\infty)}(\lambda)$ for which eq.~(\ref{formalsolutions}) are asymptotic expansions
 \begin{equation}
Y_j^{(0,\infty)}(\lambda) \sim G^{(0,\infty)} \hat Y (\lambda) e^{T^{(0,\infty)} (\lambda)} \, \ \text{in} \ \mathcal{S}_j^{(0,\infty)} \, .
\end{equation}
The solutions in overlapping sectors can differ by the constant \emph{Stokes matrices},
\begin{equation}
s_j^{(0,\infty)} := Y_{j+1}^{(0,\infty)}(\lambda)^{-1} Y_j^{(0,\infty)}(\lambda) \, ,
\end{equation}
and solutions near  $0$ and $\infty$ are related by the connection matrix $C$ 
\begin{equation}
C:= Y_{1}^{(\infty)} \left(Y_{1}^{(0)}\right)^{-1} \, .
\end{equation}
We associate to the linear system (\ref{linearsystem}) the  \emph{generalized monodromy data} $\mathcal{M}$ consisting of 
\begin{itemize}
\item Stokes matrixes $s_j^{(0,\infty)}$, $j=1,2$ 
\item Connection matrix $C$
\item ``Exponents of formal monodromy'' $T_0^{(0,\infty)}$.
\end{itemize}
This data is preserved by deformations of $T^{(0,\infty)}_{-1}$  iff a non-linear equation on the matrices $A$ of \eqref{linearsystem} in the parameter $\rho$ is satisfied.  These equations can be reduced to (\ref{PainleveIII}), i.e.\ Painlev\'e III, \cite{Jimbo:1981zz}. 
This has proved to be a powerful tool in the study of solutions to the Painlev\'e III equation. 

Although our starting point will be  twistor-theoretic, we will prove that our solutions to Painlev\'e III are characterised as those whose associated monodromy data $\mathcal{M} $ has trivial Stokes matrices and $\mathbb{Z}_2$-monodromy.

\subsection{The axis-simple condition and meromorphicity}

In Section~\ref{sec:Proof} 
we will show that the axis-simple condition on $P_{III}$ leads to the following Riemann-Hilbert problem:

\begin{thm}
\label{MainTheorem}
A solution to the $D_6$ $P_{III}$ equation is axis-simple iff it arises from the Riemann-Hilbert problem on the Riemann sphere parametrized by $\lambda$ given by  the following matrix
\begin{equation}
P = 
\begin{pmatrix}
(\rho \lambda)^{\tilde {p }} & 0 \\
0 & (\rho \lambda)^{-\tilde {p }} 
\end{pmatrix}
\begin{pmatrix}
e^{ ({a } - \tilde{a}) u } c_0 & e^{ -({a } +\tilde{a})  u } \tilde c_1 \\
 e^{ ({a } + \tilde{a})  u }  c_1 &  e^{ -({a } - \tilde{a})  u }\tilde c_0 
\end{pmatrix}
\begin{pmatrix}
(\frac{\lambda}{\rho})^{{p}} & 0 \\
0 & (\frac{\lambda}{\rho})^{-{p}} 
\end{pmatrix},
\label{MeromorphicPIII}
\end{equation}
where $(c_0,c_1,\tilde c_0,\tilde c_1,{a },\tilde{a})$ are constants with $c_0\tilde c_0-c_1\tilde c_1=1$ and ${p },\tilde{p}\in\Z$ or\footnote{In this half-integral case, we need to observe that there is no overall square-root singularity in $P$ as that in the first factor cancels that in the last.} ${p },\tilde{p}\in\Z+1/2$, and $u=\rho/2(\lambda+1/\lambda)$.  The data is gauge equivalent under pre and post multiplying by diagonal constant unit determinant matrices so that there is only one essential parameter in the $c_0,c_1,\tilde c_0,\tilde c_1$. The data is related to the standard constants by
\begin{equation*}
k^2 = \frac{1}{16} {a }^2,\ l^2 =  \tilde{a}^2,\ m=-\frac{{a }}{2}{p },\ n= -2 \tilde{a} \tilde{p},
\end{equation*}
In particular, we have the quantization condition
\begin{equation*}
\frac{1}{2} \frac{m}{k} , \frac{1}{2}  \frac{ n}{ l} \in \mathbb{Z} \mbox{ or } \Z+1/2\, .
\end{equation*}
A similar Riemann-Hilbert problem for the type  $Q$ Painlev\'e III is spelled out in the proof, whereas $D_7$ and $D_8$ are ruled out by our condition.
\end{thm}

We have from Theorem~\ref{MainTheorem} that, because the Riemann-Hilbert problem extends smoothly to $\rho=0$, we can have at worst poles in the solutions at $\rho=0$ when  the Riemann-Hilbert problem jumps there so
we have
\begin{coroll}\label{Corollary}
The axis-simple solutions presented by the Riemann-Hilbert problem in Theorem~\ref{MainTheorem} are meromorphic at $\rho=0$.
\end{coroll}

Finally, we show that the axis-simple condition can be understood from the point of view of generalized monodromy data by
\begin{thm}\label{MonodromyTheorem}
The axis-simple condition is equivalent to imposing triviality of the Stokes matrices and half-integer  exponents of formal monodromy. (We take half-integral to include integer values also.)
\end{thm}
The proof of Theorem~\ref{MainTheorem} follows by imposing the axis-simple condition on twistor space as given in \cite{Mason:1988} to reduce the twistor data to a normal form that gives rise to the above Riemann-Hilbert problem. Corollary~\ref{Corollary} follows from \cite{Mason:1988} and the proof of the Painlev\'e property in \cite{Ward:1984gw}. In this paper, we give an explicit proof for a special case. The axis-simple condition itself is reviewed in the next section.

\section{The Joukowski correspondence}

\label{sec:JukowskiCorrespondence}

In the context of the stationary axisymmetric systems, we introduce  spatial coordinates $(\rho,z )$  with $\rho$ being the radial distance from the $z$-axis. We will work on a region\footnote{Although the solutions that we are interested in are initially defined only for real $(\rho,z)$, we will be able to take $U$ to be a region in the complexification as solutions will generically be analytic in our context being solutions to elliptic equations.} $U$ in the $(\rho,z)$-plane  that we will take to be connected, simply connected,  containing some piece of the \emph{axis} $\rho=0$ and invariant under $\rho\rightarrow -\rho$.  

This requirement that the domain $U$ contains a piece of the axis $\rho=0$ is the key feature of the \emph{axis-simple case}. The corresponding solutions coming from the twistor correspondence will be allowed to have poles on the axis, but as explained below Theorem~\ref{theorem2} only as a consequence of so-called jumping lines, and therefore it will be meromorphic, but not branching. See ref.~\cite{Mason:1988} for further discussion.

\subsection{The underlying geometry}
We are defining  the \emph{Joukowski correspondence} to be  the reduced twistor correspondence introduced in \cite{Mason:1988} based on \cite{Ward:1982bf}. It is a symmetry reduction of the twistor correspondence between points in (complexified) Minkowski space-time and lines in $\CP^3$ under a time translation and spatial rotation.  It can be summarized in the double fibration
\begin{equation}
\begin{tikzcd}
& \arrow[ld, "p" above] U \times \CP^1 \ni (\rho,z, \lambda) \arrow["q" above, rd] & \\
(\rho,z) \in U & & \T (U)\ni u  .
\end{tikzcd}
\label{doublefibration}
\end{equation}
The map $p$ forgets $\lambda$ whereas $q$ projects according to
the Joukowski transformation  
\begin{equation}\label{Jouk}
u= \frac{\rho}2\left(\lambda+\frac{1}{\lambda}\right)+iz\, .
\end{equation}
This gives a family of maps from $\lambda \in \CP^1\rightarrow u\in \CP^1$ that depends on $(\rho,z)$.
At fixed $(\rho,z)$, the map $\lambda \rightarrow u$ is $2:1$, branching at $u=\pm\rho +iz$.  The unit circle $|\lambda|=1$ is mapped to the slit $[-\rho,\rho]+iz$.  
	
\begin{defn}
We define $\T(U)$ to be the space of connected leaves in $U\times \CP^1$ on which $u$ is constant. 
\end{defn}
 Na\"ively one might expect  $\T (U)$ to be the $u$-Riemann sphere and clearly there is a map $u:\T(U)\rightarrow \CP^1$ defined by $u$.  However, at a fixed value $u=u_0$, we will obtain two points when $u=u_0$ has two components in $U\times\CP^1$, although only one when the set $u=u_0$ in $U\times \CP^1$ is connected.   Given $u=u_0$,  \eqref{Jouk} gives two choices of $\lambda$ at each point in $U$ dropping to one on the branching loci $u_0=\pm\rho+iz$. Thus the criteria for $u=u_0$ in $U\times \CP^1$ to have just one component rather than two is that the branching loci $u_0=\pm \rho+iz$ should lie in $U$.  

	Identifying $U$ with a region in the $u$-Riemann sphere, we see that we have a covering $\T(U)\rightarrow \CP^1$ that is $1:1$ for $u\in U$, and $2:1$ on $\CP^1-U$.  Thus 
\begin{propn} \textnormal{\cite{Mason:1988}} $\T(U)$ is the non-Haudorff Riemann surface obtained by gluing two copies of $\CP^1$ together using the identity map on the open set $U\subset\CP^1$.
\end{propn}
In practice, we will only be concerned with the example in which $U=\C$. Then $\T(U) $ is essentially the Riemann sphere, but with two points at $\infty$.  It is  obtained by gluing two copies of the Riemann sphere $\CP^1$ together for finite values of $u$. 

Points $(\rho,z) \in U$ correspond to surjective maps 
$L_{(\rho,z)}:\CP^1 \rightarrow\T(U)$ 
given by 
$
L_{(\rho,z)}(\lambda)=q(\rho,z,\lambda)
%\frac\rho2 \left( \lambda+\frac1\lambda\right) +iz\, ,
$.

\subsection{The reduced Ward construction}

The main result that we will use follows \cite{Mason:1988} based on Ward \cite{Ward:1977ta,Ward:1982bf} concerning solutions to the stationary axisymmetric SDYM equations.  These can be expressed in the form of Yang's equation
\begin{equation}
\p_\rho(\rho J^{-1}\p_\rho J)+\p_z(\rho J^{-1}\p_z J)=0\, .
\label{SDYM-J}
\end{equation}
where $J(\rho,z)$, the \emph{$J$-matrix}, takes values in $GL(N,\C)$ in the first instance, but it can be adapted to any real or complex Lie group. For a unitary group, for example, it can be taken to be hermitian.  
\begin{thm}\label{theorem2}
There is a 1:1 correspondence between solutions to the stationary axisymmetric self-dual Yang-Mills equations on $U$ with gauge group $SL(N,\C)$ and holomorphic vector bundles $E\rightarrow \T(U)$ with structure group\footnote{Up to the caveat that at the cost of factoring out a determinant, the condition on the structure group can be relaxed. See the remark at the end of this section.}  $SL(N,\C)$ such that for any fixed $\rho+iz \in U$, the restriction of $E$ to the line $L_{(\rho,z)}={q \circ p^{-1}  (\rho+iz)} \subset \T(U)$ is trivial.
\end{thm}

\noindent
{\bf Remark:}
Although that last restriction might seem very strong,  if true at one value of $\rho+iz$, it will be true for $\rho+iz$ on a dense open subset of $U$; the points at which  it fails  correspond precisely to $J$ becoming meromorphic rather than simply holomorphic on $U$.  The points $(\rho,z)$ where $J$ is meromorphic correspond to \emph{jumping lines} $L_{(\rho,z)}$ for the bundle $E$ where it is no longer trivial, but a direct sum of nontrivial line bundles as allowed by Grothendieck's theorem.

\smallskip

 We give an outline of the proof to provide ingredients that will also be needed later. A key role is played by the vector fields 
\begin{equation}
\begin{aligned}
V_1&= \p_\rho + i \lambda \p_z + \frac{1}{\rho}\lambda \p_\lambda  
% s - \lambda (J^{-1}J_y)s &=0 
\\
V_2 &= i\p_z +\lambda \p_\rho - \frac{1}{\rho} \lambda^2 \p_\lambda \, .
% s + \lambda (J^{-1}J_r) s &=0, \, .
\end{aligned}
\label{laxvector}
\end{equation}
They satisfy $V_1u=V_2u=0$ so their integral surfaces  define the leaves of constant $u$ in $U\times \CP^1$.

 We can take the bundle $E\rightarrow \T(U)$ to be defined in the \v Cech fashion by patching functions $P(u)_{ij}$ defined on overlaps $U_i\cap U_j$ of some open cover $\{U_i\}$ of $\T(U)$. The pull-back $q^*E$ of $E$ to the Riemann sphere  $L_{(\rho, z)}$ in $U\times \CP^1$   has patching functions $P ( \rho/2 (\lambda + 1/ \lambda) + iz )_{ij}$.
 Since the bundle is assumed to be trivial on $L_{(\rho,z)}$, for fixed $(\rho, z)$ we can find $G_i$ such that
\begin{equation}
P ( \rho/2 (\lambda + 1/ \lambda) + iz )_{ij} = G_i (\rho, z, \lambda) G_j^{-1} (\rho, z, \lambda)
\label{Jfactorization}
\end{equation}
where $G_i$ is holomorphic in $\lambda$ on $q^{-1}(U_i)$.  We will normalize the solutions $G_i$ to \eqref{Jfactorization} by requiring $G_0(\rho,z,0)=1$ where $q^{-1}(U_0)$ contains $\lambda=0$ and with this, the $G_i$ are unique.    It follows from $V_1 P_{ij} = V_2 P_{ij} = 0$ that $G_i^{-1}V_1G_i=G_j^{-1}V_1G_j$ by differentiation of \eqref{Jfactorization} so that this expression is global on the $\lambda$-Riemann sphere, but with a simple pole at $\lambda=\infty$.  It follows that we can define the Lax pair by
\begin{equation}
\begin{aligned}
L_1 &:=G_i^{-1} V_1 \circ G_i= V_1  + i \lambda (J^{-1}\parz J) \\
L_2 &:=G_i^{-1} V_2 \circ G_i= V_2  + \lambda (J^{-1}\parrh J) 
\end{aligned}
\label{laxpair}
\end{equation}
where the $J$-matrix is defined by
$$
J (\rho , \lambda) = G_\infty (\rho, z, \infty) G_0 (\rho, z, 0)^{-1} \, ,$$
where $U_\infty$, $U_0$ denotes the sets whose preimage under $q$ contains $\lambda=\infty, 0$ respectively.   The $L_1$ and $L_2$ simultaneously annihilate $G_i^{-1}$ by construction. Thus, the Lax pair is compatible
\begin{equation}
[L_1,L_2]=0. 
\label{LaxCompatibility}
\end{equation}
This is equivalent to Yang's equation form of the axisymmetric self-dual Yang-Mills equations \eqref{SDYM-J}.

\smallskip

The significance of the theorem is that  holomorphic vector bundles on $\T(U)$ can be described in terms of essentially free data $P_{ij}(U)$. although subject to certain consistency conditions.  However, the trivialization problem for the bundle is still complicated as there are potentially many sets in an open cover, and the presentation  is still subject to a large degree of redundancy also corresponding to changes of frames on the open sets.  We will see that it can be put into a standard normal form in interesting  situations.

\subsection{The normal form in the axis-simple case}

In this section, we introduce a normal form for bundles on twistor space satsifying the axis-simple condition.  The task of constructing solutions will then be reduced to the Riemann-Hilbert problem eq.~(\ref{Jfactorization}) based on
the patching data $P$ in this normal form.  This section is essentially a review of Section~5.4 of \cite{Mason:1988} and concerns axis-simple stationary axisymmetric solutions to the self-dual Yang-Mills equations.  In Section~\ref{sec:InvariantBundles} we reduce the description to one for Painlev\'e III.

\begin{propn}
In the axis-simple case, a vector bundle over $\T (U)$ of rank $N$ and structure group $SL(N, \C) $ is characterized completely by a set of $2N-2$ integers $(p_1,\ldots,p_N,\tilde p_1,\ldots ,\tilde p_N)$, $\sum_ip_i=\sum_i\tilde p_i=0$ and a holomorphic matrix $P(u)$ on $U$ with values in $SL(N,\C)$. 

The reconstruction of the $J$-matrix  at $(\rho,z)$ arises from a Riemann-Hilbert problem on the $L_{(\rho,z)}$ Riemann sphere parametrized by $\lambda$ given by 
\begin{equation}
\widetilde P(\lambda, \rho,z ) = G_{\infty} (\rho,z,\lambda) G_{0} (\rho,z,\lambda)^{-1} .
\label{RH}
\end{equation}
where
\begin{equation}
\widetilde P  =  
\begin{pmatrix}
(\lambda \rho)^{\tilde p_1}  &  & 0 \\
& \ddots &  \\
0 &  & ( \lambda \rho)^{\tilde p_N}
\end{pmatrix} P_U (u)
\begin{pmatrix}
\left( \lambda / \rho  \right)^{ p_1} &  & 0 \\
 & \ddots &  \\
0 & &  \left(  \lambda / \rho \right)^{ p _N}
\end{pmatrix} .
\label{patching}
\end{equation}
We then have $J(\rho,z) =G_\infty(\rho,z,\infty)G_0(\rho,z ,0)^{-1}$\, .
\end{propn}

The key simplification in the axis-simple case is that, as me mentined, $\T (U)$ consists of two copies of the Riemann sphere glued together over $U$. We use the \v{C}ech description in terms of patching matrices to describe bundles  $E \rightarrow \T (U)$. Let the rank of $E$ be $N$, and denote the two Riemann spheres by $\CP^1_0$ (that contains $\lambda=0$ in its pre-image) and $\CP^1_\infty$ that contains $\lambda=\infty$. We cover $\T (U)$ with four open sets: $U_0$ a copy of $U$ in $\CP^1_0$, $V_0$ a neighbourhood of $u=\infty \in \CP^1_0$ intersecting $U_0$ in an annular region, and two analogous open sets $U_\infty$ and $V_\infty$ covering $\CP^1_\infty$. 
A standard theorem due to Birkhoff and Grothendieck gives that the restriction of $E$ to either sphere must be a direct sum of line bundles:
$$
E|_{\CP^1_0} \cong \bigoplus_{i=1}^N \mathcal{O}(p_i), \quad  E|_{\CP^1_\infty}  \cong \bigoplus_{i=1}^N \mathcal{O}(\tilde p_i), \quad p_i \in \mathbb{Z}, \ \tilde p_i \in \mathbb{Z} 
$$
where the $p_i$s and $q_i$s are the Chern classes of the line bundles. Since $(U_0,V_0)$ and $(U_\infty, V_\infty)$ are standard covers of the spheres, it follows that we can choose holomorphic frames on the four sets such that the transition matrices on $U_0\cap V_0$ and $U_\infty\cap V_\infty$ are of the form $\mathrm{diag}( u^{p_1}, \cdots u^{p_N} )$ and $\mathrm{diag} ( u^{\tilde p_1}, \cdots u^{\tilde p_N} )$ respectively. Assuming the structure group to be $SL (N, \C)$, we also have the constraints $\sum_{i=1}^{N} p_i = \sum_{i=1}^{N} \tilde p_i = 0$. Lastly, the frames on $U_0$ and $U_\infty$ must be patched by an undetermined $N \times N$ matrix of unit determinant, which we call $P_U$:
$$
P_U : U_0 \cap U_\infty= U \ \rightarrow SL (N, \C) .
$$
Thus, a vector bundle over $\T (U)$ of rank $N$ and structure group $SL(N, \C) $ is characterized as described in the proposition above in the axis-simple case.

We can reduce the four set cover Riemann-Hilbert problem of \eqref{Jfactorization} to a two-set one on $L_{(\rho,z)}$ in terms of  $\lambda$ as follows.  Because the patching matrices on $U_0\cap V_0$ and $U_{\infty} \cap V_{\infty}$ are  particularly simple, after pulling them back to $L_{(\rho,z)}$, \emph{i.e.} after the substitution $u=\rho/2(\lambda+1/\lambda)+iz$, we can factorize $u$ and hence its powers by 
\begin{equation}
u=\frac{\rho\lambda}2 \times \left( 1+\frac{1}{\lambda^2}+\frac{2iz}{\rho\lambda} \right)\quad \mbox{near $\lambda=\infty$, } \label{scal-RH}
\end{equation}
and 
\begin{equation}
 u=\frac\rho{2\lambda}\times  \left( 1+\lambda^2+\frac{2iz\lambda}\rho \right) \quad\mbox{ near }\lambda=0\, . \label{scal-RH0}
\end{equation}
We use the second factors to find  frames of the pullback of $E$ to $L_{(\rho,z)}$ on  the two sets $\widetilde U_0=\{|\lambda| < 2\}$ and $\widetilde{U}_\infty =\{|\lambda|>1/2\}$.  In these frames, there is now just the one patching matrix on $\widetilde U_0\cap \widetilde U_\infty$ as given in \eqref{patching}. 

The last step to find the global frame required in \eqref{Jfactorization} is to solve a  Riemann-Hilbert problem: 
given an invertible holomorphic matrix $\widetilde P(\lambda) \in SL(N, \C)$ defined in a neighbourhood of $|\lambda|=1$ (an element of the loop group $LSL(N, \C)$), we look for $SL(N,\C)$-valued functions $G_{\infty} (\lambda )$ and $G_0 (\lambda )$ holomorphic in $\lambda$ for $|\lambda| > 1-\epsilon $ and $|\lambda|< 1+ \epsilon$ for $\epsilon \in \mathbb{R}^{+}$ such that
\begin{equation}
\widetilde P(\lambda ) = G_{\infty} (\lambda) G_{0} (\lambda)^{-1} .
\end{equation}
According to Birkhoff's theorem, for generic loop group elements $P(\lambda)$ solutions exist\footnote{More precisely, ``generic" here means that the element is in the identity component of $LSL(N,\C)$ endowed with uniform convergence topology. The ``non-generic'' elements have an additional diagonal contribution, which corresponds precisely to the diagonal transition matrices in Birkhoff-Groethendieck's theorem. Thus, ``generically" any holomorphic bundle on $\mathbb{C}\mathbb{P}^1$ is trivial.} and are unique up to multiplication by a constant matrix $C$,
\begin{equation}
G_{\infty} \mapsto G_{\infty} C, \quad G_0 \mapsto G_0 C .
\end{equation}
In our context, we have a family of $P(\lambda)$s parametrized by $(\rho,z)$, and $C$ can then depend on $(\rho,z)$. Although the transormation leaves $J$ invariant, we remark that it actually corresponds to a gauge transformation of the SDYM connection associated with $J$.  
\\

\noindent
{\bf Remark:} 
we will  relax the $SL(N,\C)$ condition on the bundles mildly so that $\sum_{i=1}^{N} p_i + \tilde p_i = 0$. This condition leads to $SL(N, \C)$ solutions to the Ernst equation up to a determinant factor consisting of powers of $\rho$ that can be removed.\\

Although Birkhoff's factorization theorem gives a generic existence theorem for solutions to the Riemann-Hilbert problem, to obtain explicit solutions, we need to make some further restrictions on the data as will be  decribed in Section~\ref{sec:WardAnsatz}.

 \section{Meromorphic Painlev\'e III transcendants}
 \label{sec:Proof}

In this section we adapt the above construction to Painlev\'e III and derive Theorem~\ref{MainTheorem}. First, we study the Lax pair for $P_{III}$ and then go on to characterize the axis-simple holomorphic vector bundles on $\mathbb{T}(U)$ that are invariant under the action of the translational Killing vector $\p_z$. These are the bundles that yield $P_{III}$ solutions. We finally map the free parameters entering the vector bundles to the constants $k$, $l$, $m$, $n$ parametrizing the type of $P_{III}$. 
\subsection{The Lax pair and isomonodromy}

The Lax pair for  $P_{III}$ arises from that  for the stationary axisymmetri Yang-Mills equations eq.~(\ref{laxpair}) when the fields are independent of $z$. However, we have tacitly made a gauge choice when writing eq.~(\ref{laxpair}), that does not allow $z$-independent fields and so is unsuitable for deriving $P_{III}$. We therefore write the Lax pair in a general gauge as 
\begin{align}
L_1 &  = \parrh + \frac{\lambda}{\rho} \partial_{\lambda} - A - \lambda B 
\nonumber %\label{LaxPain1}
 \\
L_2 & =   \lambda \parrh - \frac{\lambda^{2}}{\rho} \partial_{\lambda} -  C -\lambda D, \label{LaxPain2}
\end{align}
where the $\mathfrak{s}\mathfrak{l}(2, \mathbb{C})$-valued functions $A$, $B$, $C$, $D$ depend only on $\rho$. We then require as a compatibility condition that the Lax pair commutes up to a linear combinations of itself.
After making the gauge choice $A=0$, this implies
\begin{equation}
\begin{aligned}
&\parrh C = 0, \\
& \rho \parrh B = [\rho D,  B], \\
&\parrh\left( \rho D \right) = \rho [B,C].
\end{aligned}
\label{PIIIsystem}
\end{equation}
The derivation of $P_{III}$ from this system is given in full detail in\footnote{
The system in \cite{Mason:1996} is given by
$%\begin{equation}
%\begin{aligned} &
\parrh P = 0, %\qquad %\\ & 
\parrh Q = 2 [Q,  R], %\qquad %\\ &
\parrh R = 2 \rho [Q,P]
%,\end{aligned}
$
\ and comparing with eq.~(\ref{PIIIsystem}) for $A$, $B$, $C$, $D$,  
we find 
$R= -\frac{1}{2} \rho D , \quad Q=B, \quad P =-\frac{1}{4} C $.} \cite{Mason:1996}, p.103. 
We wish to relate the Painlev\'e transcendent $f$ and the constant parameters $(\alpha , \beta, \gamma , \delta)$ to the matrices $A$, $B$, $C$, $D$ entering the Lax pair. 
Comparing with\footnote{
The constants of motion are given in terms of $P$, $Q$ and $R$ by $
k^2 = \frac{1}{2} \tr (P^2) = \frac{1}{32} \tr (C^2), l^2 = \frac{1}{2} \tr (Q^2) = \frac{1}{2} \tr (B^2), 
m= \tr (PR) = \frac{\rho}{8} \tr (CD), \quad n = \tr (QR) = \frac{\rho}{2} \tr (BD) .
$} \cite{Mason:1996}, we find that  the constants are given by
\begin{alignat}{3}
k^2 &= \frac{1}{32} \tr ( C^2 ), \quad &m& = \frac{\rho}{8} \tr (C D), \nonumber \\
l^2 &=  \frac{1}{2}\tr (B^2), \quad &n& = -\frac{\rho}{2} \tr (B D), \label{constants}
\end{alignat}
where $k$, $l$, $m$, $n$ are the parameters introduced in \eqref{parameterchange}. The transcendent reads
\begin{equation}
f=
\begin{cases}
-\frac{1}{2} D_{12} / B_{12} & \text{if} \ k \neq 0  \\
 -\frac{1}{2} D_{21} / B_{21} & \text{if} \ k = 0 \, ,
\end{cases}
\label{transcendent}
\end{equation}
where the subscripts indicate the respective entries of the matrices in a frame in which $C$ is diagonalized when $k\neq 0$, or is strictly upper triangular for $k=0$.

\paragraph{The monodromy operator:} eliminating $\p_\rho$ from \eqref{LaxPain2} we obtain  
\begin{equation}
\partial_{\lambda} -\mathcal{A}(\lambda) :=  \p_\lambda + \rho  \frac{D}{2 \lambda}  +\rho \frac{  C}{2 \lambda^2} - \rho \frac{  B}{2}  .
\label{isomonodromy}
\end{equation}
This defines a holomorphic flat connection on the Riemann sphere with double poles at $\lambda=0$ and $\infty$ that defines the isomonodromy problem associated with the linear system eq.~(\ref{isomonodromy}).  The compatibility with \eqref{LaxPain2} means that the generalized monodromy of the opeator is independent of $\rho$. 

We will use this flat connection to express $B$, $C$, $D$, in terms of the geometric data representing solutions to $P_{III}$.

\subsection{Characterization of invariant bundles}
\label{sec:InvariantBundles}

By differentiation of the incidence relation $u = \frac{\rho}2 (\lambda + 1/\lambda) + iz$ it is easy to see that the symmetry  $\p_z$ acts by $\p_u$ on $\mathbb{T}(U)$.  This has $u = \infty$ as a fixed point, so we cannot  simply quotient the space by this action to construct invariant bundles as pullbacks from a quotient.  Instead, we must characterise vector bundles $E\rightarrow\mathbb{T}(U)$ that carry a global holomorphic lift $\Lie_{\p_u}$ of $\p_{u}$.   In the following, we study axis-simple $GL(2,\C)$ bundles, and so without loss of generality
$$ E|_{\CP^1_0} \cong \CO ({p }) \oplus \CO (q), \quad E |_{\CP^1_\infty} \cong \CO (\tilde{p}) \oplus \CO (\tilde q) .$$
Since the following discussion applies equally to both spheres, for this first part we will drop the subscripts $\{0,\infty\}$, and refer generically to ``the sphere". We work locally on the sphere's copy of $U$, and assume that a frame has been chosen so that the matrix patching $U$ with a neighborhood $V$ of $\infty \in \CP^1$ is of standard form, 
\begin{equation}
P_{UV} = \mathrm{diag}(u^{-p},u^{-q}).
\label{spherepatching}
\end{equation}

\paragraph{Generic case.} Assume 
that $p > q$. 
 The action of $ \p_u$ on $E$ is expressed in terms of a Lie derivative $\mathcal{L}_{\p_u}$. Locally on $U$, we have
\begin{equation}
\mathcal{L}_{\p_u} =  \p_u + \theta_U ,\quad \mbox{ where } \quad 
\theta_U =\begin{pmatrix}
a&d\\c&b
\end{pmatrix},\label{Lie-def}
\end{equation}
and $\theta_U$ must be holomorphic on $U$. On $V$ we will therefore have
$$
\mathcal{L}_{\p_u} = P_{UV}^{-1}(\p_u+\theta_U)P_{UV}= \p_u + \theta_V =\p_u + P_{UV}^{-1}(\p_u P_{UV}) + P_{UV}^{-1}\theta_UP_{UV},$$
so 
\begin{equation}
\theta_V=\begin{pmatrix}
a -\frac{p}{u} &du^{p-q}\\
cu^{q-p}& b-\frac{q}{u}
\end{pmatrix} \label{theta-infty}
\end{equation}
and this must be holomorphic near $\infty$.  Since $p>q$ this implies $d=0$, $a$ and $b$ are constants, and $c$ is a polynomial in $u$ of degree at most $p-q$, i.e.\
\begin{equation}
 d=0, \qquad a,  b\in \mathbb{C}, \qquad  c(u)= \sum_{k=0}^{p-q} c_k u^k  
\end{equation}

 The general form of $\mathcal{L}_{\p_u}$ can be restricted further by making use of the residual gauge transformations that preserve the patching matrix eq.~(\ref{spherepatching}). These are given by
\begin{equation}
G=
\begin{pmatrix}
e & 0 \\
f & g 
\end{pmatrix},
\end{equation}
where again $k$, $m$ are constants, whereas $l$ is a polynomial of degree $p-q$. Applying such a gauge transformation to $\theta$ does not change $a$ or $d$, but gives
\begin{equation}
 c'=
\frac{ce+ (b-a)f +\p_u l}{g}\, .
%\frac{i}{m}\sum_{k = 0}^{p-q-1} \left( l_{k+1} + gc_k + (b-a) l_k \right)u^{k} + \frac{1}{m} \left( g c_{p-q} + (b-a) l_{p-q} \right)u^{p-q}.
\label{remainder}
\end{equation}
We assume for the moment $a\neq b$, and also include this (besides the assumption on the Chern classes) as a condition for the \emph{generic} case. In this generic case,  $l$ can be chosen to cancel $c'$ precisely.  Thus, without loss of generality, we can take 
\begin{equation}
\mathcal{L}_{\p_u} = \paru + \theta =  \paru +
\begin{pmatrix}
a & 0 \\
0  & b
\end{pmatrix}.
\label{StandardLie}
\end{equation}
Note that we did not make use of $e$, $g$ so that we still have a remaining diagonal gauge freedom 
\begin{equation}
G=
\begin{pmatrix}
e & 0 \\
0 & g
\end{pmatrix}.
\label{diagonalgauge}
\end{equation}

Having put the Lie derivative in standard form on both spheres, we now study how they are related to each other on $U$ and reintroduce subscripts $\{0,\infty\}$ to distinguish between them. The crucial point is that the patching matrix $P_U$ must send the Lie derivative defined on $U_0$ to that defined on $U_{\infty}$. This implies that 
\begin{equation}
 \paru P_U  = P_U \theta_{\infty} -  \theta_{0} P_U,
\label{PODE}
\end{equation}
which is a first order matrix ODE for $P_U$. The solution is easily written in terms of exponentials
\begin{equation}
P_U = \exp \left(-  \theta_{\infty}  u \right)  C \exp \left(  \theta_{0} u \right),
\end{equation}
where $C$ is an arbitrary invertible matrix with constant entries, say $c_0$, $c_1$, $\tilde c_0$, $\tilde c_1$. With $\theta_0$ and $\theta_\infty$ as in eq.~(\ref{StandardLie}),
\begin{equation}
P_U = 
\begin{pmatrix}
e^{ ({a } - \tilde{a})  u } c_0 & e^{ (b - \tilde{a})  u } \tilde c_1\\
 e^{ ({a } - \tilde b) u } c_1 &  e^{ (b - \tilde b)  u } \tilde c_0
\end{pmatrix}.
\label{Psolution}
\end{equation}
The solution still contains some redundancy, because of the residual diagonal gauge freedom   eq.~(\ref{diagonalgauge}). This means that we are free to multiply $P_U$ from the left and from the right by two different constant diagonal matrices. As a consequence, out of the four $c$s, only one is essential. 

This completes the characterization of $\p_u$ invariant vector bundles in the generic case.  It follows from the previous section that this data leads to the patching matrix
\begin{equation}
P = 
\begin{pmatrix}
(\rho \lambda)^{\tilde {p }} & 0 \\
0 & (\rho \lambda)^{ \tilde q} 
\end{pmatrix}
\begin{pmatrix}
e^{ ({a } - \tilde{a})  u } c_0 & e^{ (b - \tilde{a})  u } \tilde c_1 \\
 e^{ ({a } - \tilde b)  u } c_1 &  e^{ (b - \tilde b)  u } \tilde c_0
\end{pmatrix}
\begin{pmatrix}
(\frac{\lambda}{\rho})^{{p}} & 0 \\
0 & (\frac{\lambda}{\rho})^{ q} 
\end{pmatrix}\, .
\end{equation}

\paragraph{Reduction to $\mathrm{SL}(2,\mathbb{C})$.} For Painlev\'e equations we only need $\mathrm{SL}(2,\mathbb{C})$-bundles. First of all, as explained above, in this case we must have ${p }+q = \tilde{p} + \tilde q$. We can shift  ${p }+q$ to zero by multiplying by a multiple of the identity, perhaps at the cost of introducing a half-integer value for ${p }$ and $q$ when ${p }+q$ is odd (and hence also of $\tilde{p}$ and $\tilde q$) leading to the condition 
$$
{p }, \tilde{p} \in \Z \mbox{ or } \Z+1/2 \, .
$$
Since elements of $\mathfrak{sl}(2, \mathbb{C})$ are traceless, we have ${a } = -b$, $\tilde{a} = -\tilde b$. The presentation of the bundle becomes
\begin{equation}
P = 
\begin{pmatrix}
(\rho \lambda)^{\tilde {p }} & 0 \\
0 & (\rho \lambda)^{- \tilde {p }}
\end{pmatrix}
\begin{pmatrix}
e^{ ({a } - \tilde{a}) u } c_0& e^{ (-{a } - \tilde{a}) u } \tilde c_1\\
 e^{ ({a } + \tilde{a}) u } c_1 &  e^{ (-{a } + \tilde{a}) u } \tilde c_0
\end{pmatrix}
\begin{pmatrix}
(\frac{\lambda}{\rho})^{{p}} & 0 \\
0 & (\frac{\lambda}{\rho})^{-{p}}
\end{pmatrix},
\label{SL2Solution}
\end{equation}
where $c_0\tilde c_0-c_1\tilde c_1=1$ and the unit determinant diagonal gauge transformations reduce the $c$s  to one essential parameter. This is eq.~(\ref{MeromorphicPIII}) in the theorem above. Lastly, notice that the degenerate case can only arise when ${a } = 0$ or $\tilde{a}=0$.

\paragraph{Non-generic cases.} We relax the generic restrictions that we have put above. First assume ${a }=b$ in $\theta_0$. Sticking to the $\mathrm{SL}(2,\C)$ case, we must have ${a }=b=0$.  Now in eq.~(\ref{remainder}) we cannot cancel the leading term of the polynomial $\tilde c$. Thus we arrive at the standard form 
\begin{equation}
\mathcal{L}_{\p_u} = \paru + \begin{pmatrix}
0 & 0 \\
cu^{2p} & 0
\end{pmatrix}\, ,
\label{degenerateLieDerivative}
\end{equation}
where now $c$ is a constant. There are no substantial differences in the subsequent discussion. The ODE eq.~(\ref{PODE}) can be solved in terms of exponentials, but we need to take into account that one of the $\theta$s depends on $u$. Assuming \emph{e.g.} that the degenerate case holds for $\theta_0$, the solution becomes
\begin{equation}
P_U = \exp \left(-\theta_{\infty}   u \right) \cdot C \cdot \exp \left(  \Theta_{0} (u)\right),
\end{equation}
where $\Theta_0 (u)$ is the primitive of $\theta_0 $,
\begin{equation}
\Theta_0  (u) = \begin{pmatrix}
0 & 0 \\
\frac{cu^{2p+1}}{2p+1}  & 0
\end{pmatrix}.
\end{equation}
The exponential of $\Theta_0 (u)$ is
\begin{equation}
\begin{split}
\exp (\Theta_0 (u))  =& \begin{pmatrix} 1 & 0 \\ \frac{cu^{2p+1}}{2p+1}  & 1 \end{pmatrix}
\end{split}.
\end{equation} 
When ${p } = q $ (and so $p=0$ in the $\mathrm{SL}(2,\mathbb{C})$ case) and/or $\tilde{p} = \tilde q$ (and similarly $\tilde{p}=0$ restricting to $\mathrm{SL}(2,\mathbb{C})$),  dropping subscripts for a moment, all entries of $\theta$ must be constant from the argument above.
Gauge transformations are now  constant matrices and simply conjugate $\theta$. Generically, $\theta$ is diagonalizable, and so we can still put $\theta$ in the standard form of eq.~(\ref{StandardLie}) by means of a gauge transformation. Otherwise, we can put the matrix into strictly lower triangular  form, and treat the nilpotent case as above.

\subsection{Identification of the parameters.} 

We now compare the data in the normal forms above to the Painlev\'e constants in the definition of $P_{III}$, namely the complex numbers $k$, $l$, $m$, $n$, and to the initial conditions. we prove :
\begin{lemma}
The Painlev\'e parameters are given in terms of the data for  invariant bundles given in the previous subsection by
\begin{itemize}
\item{In the generic case ($a \neq 0$, $\tilde a \neq 0$, $p \neq 0$, $\tilde p \neq 0$)
\begin{equation*}
k^2 = \frac{1}{16} {a }^2,\ l^2 =  \tilde{a}^2,\ m=-\frac{{a }}{2}{p },\ n= -2 \tilde{a} \tilde{p}, \, .
\end{equation*}
and the Painlev\'e equation is of the type $D_6$. }
\item{If ${p }=0$, then $m=0$ with no restriction on the other parameters, which are given by the same formulae as above. Similarly if $\tilde{p}=0$ then $n=0$, with the other parameters given as above.}
\item{If ${a }=0$, then $k=m=0$. Similarly, if $\tilde{a}=0$ then $l=m=0$. Therefore, the Painlev\'e equation is of type $Q$.}
\end{itemize}
We see in particular that the $D_7$ and $D_8$ cases do not occur. The only free parameter that can correspond to initial conditions is encoded in $c_0$, $c_1$, $\tilde c_0$, $\tilde c_1$ up to diagonal gauge transformations and subject to $c_0\tilde c_0-c_1\tilde c_1=1$. \end{lemma}

We prove the lemma by identifying our parameters in terms of the invariants of the matrices in the the isomonodromy operator $\CA(\lambda)$ in eq.~(\ref{isomonodromy}). We start by sketching how eq.~(\ref{isomonodromy}) arises from the construction from the twistor data\footnote{For more details on this point, see \cite{Mason:1996}, p.232.}. 

The basic idea is that $du\Lie_{\p_u}$ defines a flat holomorphic connection $\nabla$ on the bundle  $E\rightarrow\mathbb{T}(U)$ with a double pole at $u=\infty$ (since $du$ has a double pole and $\theta$ does not vanish).  This pulls back to give the isomonodromy operator on  the pullback  $q^*E$  of $E$ to $U\times \CP^1$ along the $\CP^1$ factor. 

The isomonodromy operator is defined on $q^*E$ along $L_{\rho , z}$ and given in \eqref{isomonodromy}  as
\begin{equation}
\nabla f =  d\lambda (\partial_{\lambda} - \CA(\lambda)) f.
\end{equation}
For $P_{III}$ this has double poles at $\lambda=0,\infty$.  The Painlev\'e constants $k,l,m,n$ are the invariants of $\CA$ at these poles defined by \eqref{isomonodromy} and \eqref{constants}.

We can obtain $\nabla$ near $\lambda=0,\infty$ from our formulae for $du\Lie_{\p_u}$ above.  First of all we must use the formula \eqref{theta-infty} valid near $u=\infty$ either on $\CP^1_0$ or $\CP^1_\infty$ for $\Lie_{\p_u}$. We work to start with in the generic $SL(2,\C)$ case and so we will have
\begin{equation}
\theta_{0V}=\begin{pmatrix}
{a }-\frac{{p }}{u}&0\\0&-{a }+\frac{{p }}u
\end{pmatrix}\, , \quad \theta_{\infty V}=\begin{pmatrix}
\tilde{a}-\frac{\tilde{p}}{u}&0\\0&-\tilde{a}+\frac{\tilde{p}}u
\end{pmatrix}
\end{equation}
where we have put the extra $0$ or $\infty$ subscript on $\theta_{V}$ to denote the version of \eqref{theta-infty} appropriate to $\CP^1_0$ or $\CP^1_\infty$.

The operator in \eqref{isomonodromy} is a holomorphic gauge transformation of $du\Lie_{\p_u}$ near $u=\infty$ obtained from the solution to the Riemann-Hilbert problems \eqref{Jfactorization} which we reduced to \eqref{RH} by means of \eqref{scal-RH0} and \eqref{scal-RH}.
The combined effect is a gauge transformation $G_0$ that is holomorphic near $\lambda=0$ (and $G_\infty$ near $\lambda=\infty$).  Thus focussing first near  $\lambda=0$
\begin{equation}
du(\p_u + \theta_{0V})=G_0^{-1}\nabla G_0 =d\lambda \p_\lambda + d\lambda (G_0^{-1}\p_\lambda G_0 +G_0^{-1} \CA G_0)\,.
\end{equation}
  Because $G_0$ is holomorphic near $\lambda=0$, the singular terms transform homogeneously. Using  $u=\rho/2\lambda +O(1)$ and $du=-\rho d\lambda/2\lambda^2$ near $\lambda=0$,
$$
G_0 \CA(\lambda)G_0^{-1}=-\frac{\rho}{2\lambda^2}\begin{pmatrix}
{a }-\frac{2\lambda {p }}{\rho}&0\\ 0& -{a } +\frac{2\lambda {p }}{\rho}
\end{pmatrix} + O(1)
$$
 Thus we have 
\begin{align}
\mathrm{tr}  \left( \CA (\lambda)^2 \right)  = \frac{\rho^2{a }^2 }{2 \lambda^4} - \frac{2\rho a p}{\lambda^3}  + O\left( \frac{1}{\lambda^2}\right) .
\end{align}
Given that
$$
G_0 \CA(\lambda)G_0^{-1}=G_0\frac\rho2 \left(\frac D\lambda +\frac C{\lambda^2}\right)G_0^{-1}+O(1) \, ,
$$ 
we can then compare the fourth-order and third-order poles, and read off the desired constants of motion:
\begin{equation}
k^2 = \frac{1}{32} \mathrm{tr} \left(C^2 \right) =   \frac{{a }^2}{16} , \quad m = \frac{\rho}{8} \mathrm{tr} \left( D C \right)  = -\frac{{a }{p }}{2}.
\end{equation}
Lower order singularities do not yield isomonodromy invariants.

Similarly, working near $\lambda=\infty$ we obtain
\begin{equation}
l^2 =\frac{1}{2} \mathrm{tr} \left( B^2 \right) = \tilde{a}^2 , \quad n = \frac{\rho}{2} \mathrm{tr} \left( B D \right) =  -2 \tilde{a}   \tilde{p} \, .
\end{equation}
We have thus mapped the constants of motion to the geometric data in the generic case, and the only free parameter left, one of $c_0$, $c_1$, $\tilde c_0$, $\tilde c_1$ encodes therefore the initial conditions.

{The degenerate cases can be treated similarly, using the respective standard form for the Lie derivative given for example in eq.~(\ref{degenerateLieDerivative}) giving the result stated in the lemma.}

\subsection{Behaviour as $\rho\rightarrow 0$}

Although Corollary~\ref{Corollary}, or equivalently meromorphicity at $\rho=0$ of the solutions encoded in Theorem~\ref{MainTheorem}, follows from the general result of \cite{Ward:1984gw} together with the axis-simple condition, in this subsection we investigate this explicitly in a particular case of $D_6$, namely ${p } = -\tilde{p}$ in eq.~(\ref{MeromorphicPIII}). We also rescale $c_0$, $c_1$, $\tilde c_0$, $\tilde c_1$ so that $c_0=\tilde c_0 = 1$, $ c_1 = -1/ \tilde c_1$. We show that singularities at the origin are simple poles and that the residue is fixed to be $-\alpha/\gamma $. This result is consistent with a standard Frobenius anaylsis\footnote{
In order to show this, we multiply eq.~(\ref{PainleveIII}) through by $f$, and insert $f=\sum_{k\geq 0} a_k \rho^{k+c}$. Assuming $c \leq -2$, we see that the lowest powers in $\rho$ give the constraints 
\begin{equation}
\gamma a_0^4 \rho^{4c} = 0 , \quad \alpha a_0^3 \rho^{3c-1} = 0 \, ,
\end{equation}
which, provided both $\gamma$ and $\alpha$ are not zero, implies that $a_0 = 0$. If $c=-1$, the lowest term gives
\begin{equation}
-( \alpha a_0^3 + \gamma a_0^4 ) \rho^{-4} = 0 \, ,
\end{equation}
which gives 
\begin{equation} \label{a_0}
a_0 = - {\alpha}/{\gamma} .
\end{equation}
Thus, unless $\alpha = \gamma = 0$, the transcendent has at most a simple pole at $\rho = 0$. 
}.

Our argument is similar to the one in \cite{Mason:1988}, p.94, and it goes in two steps. First, we perform a splitting of the patching matrix in eq.~(\ref{MeromorphicPIII}) in the $\rho \rightarrow 0$ limit, and thus obtain an expression for the $J$-matrix in this limit. Second, we express the system eq.~(\ref{PIIIsystem}) in terms of the $J$-matrix, so that we can see how the Painlev\'e transcendent is given in terms of its components.

The first step is to find $G_0(\rho,z,\lambda)$ defined for $\lambda\neq \infty$ and $G_\infty(\rho,z,\lambda)$ so that $P=G_\infty G_0^{-1}$ which we will do in a series in $\rho$.
We assume ${p } = -\tilde p$ in eq.~(\ref{MeromorphicPIII}), and for notational simplicity we define $\alpha={a } - \tilde{a} $, $\beta= {a } + \tilde{a}$. Setting $u= \frac{\rho}{2} \left( \lambda + 1/\lambda \right) + iz$ to split $e^{\alpha u}=e^{\alpha(iz+ \rho/2\lambda)}e^{\alpha\rho\lambda/2}$ we have
\begin{equation}
\begin{split}
P&=\frac{1}{\sqrt{2}} 
\begin{pmatrix}
\rho^{\tilde{p}} & 0 \\
0 & \rho^{-\tilde{p}}
\end{pmatrix}
\begin{pmatrix}
e^{ \alpha u }  & \lambda^{2\tilde{p}}e^{ \beta u }  \\
-\lambda^{-2\tilde{p} } e^{ -\beta u }  &   e^{ - \alpha u } 
\end{pmatrix}
\begin{pmatrix}
\rho^{\tilde{p}} & 0 \\
0 & \rho^{-\tilde{p}}
\end{pmatrix}  \\
&=
\frac{1}{\sqrt{2}}\begin{pmatrix}
\rho^{\tilde{p}} e^{\alpha(iz+\rho/2\lambda)} & 0 \\
0 & \rho^{-\tilde{p}} e^{-\alpha(iz+\rho/ 2\lambda)}
\end{pmatrix} \\
&\times \begin{pmatrix}
1 & c_1 \lambda^{2\tilde{p}}e^{ \beta u -\alpha \rho( 1/ \lambda - \lambda)/2  - \alpha i z }  \\
-1/c_1\lambda^{-2\tilde{p}} e^{ - \beta u +\alpha \rho ( 1/ \lambda - \lambda)/2  + \alpha i z }  &  1 
\end{pmatrix} \\
& \times
\begin{pmatrix}
\rho^{\tilde{p}}e^{\alpha\rho\lambda/2  } & 0 \\
0 & \rho^{-\tilde{p}}e^{-\alpha\rho\lambda/2  }
\end{pmatrix} \, .
\end{split}
\end{equation}
Without further loss of generality, we assume
$\tilde{p}\geq 0$, and we focus on the matrix in the middle of the RHS, which we call $\tilde{P}$. Its off diagonal entries can be expanded in $\rho$ and if we do so up to $\rho^{2\tilde{p}}$ 
\begin{equation}
\tilde{P}=\
\begin{pmatrix}
1 & g  \\
-\hat{g}  &  1
\end{pmatrix} 
+
 \mathcal{O}\left(\rho^{2\tilde{p}+1}\right) \, ,
\end{equation}
we find that $g$ is a polynomial in $\lambda$
$$
g= c_1 e^{i(\beta-\alpha)z}\lambda^{2\tilde{p}}\sum_{i=0}^{2\tilde{p}} \frac{\rho^i((\alpha+\beta)\lambda +(\alpha-\beta)/\lambda)^i}{2^i i!}= \sum_{i=0}^{4{\tilde{p} }} g_i \lambda^{i},  
$$
where 
$$
 g_0 =c_1 \frac{((\alpha-\beta)\rho)^{2\tilde{p}}e^{i(\beta-\alpha)z}}{2^{2\tilde{p}}(2\tilde{p})!} 
$$
is generically nonvanishing.  There is
a similar formula for $\hat g$ as a polynomial in $1/\lambda$, $\hat{g}= \sum_{i=0}^{4{\tilde{p} }} \hat{g}_i \lambda^{-i}$, with $\hat{g}_0$ generically nonvanishing.  Notice further that from their definitions
\begin{equation}
g \hat{g} = 1 + \mathcal{O}\left( \rho^{2\tilde{p}+1} \right) \, .
\end{equation}
We can now easily split the matrix $\tilde{P}$ up to order $\mathcal{O}\left( \rho^{2\tilde{p}+1} \right)$,
\begin{equation}
\begin{split}
\tilde{P}&=
\begin{pmatrix}
1 & 0  \\
-\hat{g}  &  1 + g \hat{g} 
\end{pmatrix} 
\begin{pmatrix}
1 & g \\
0 & 1
\end{pmatrix}
+
\mathcal{O}\left(\rho^{2\tilde{p}+1} \right)  \\
&=\begin{pmatrix}
1 & 0  \\
-\hat{g}  &  2 
\end{pmatrix} 
\begin{pmatrix}
1 & g \\
0 & 1
\end{pmatrix}
+
\mathcal{O}\left(\rho^{2\tilde{p}+1} \right) 
\end{split} \, .
\end{equation}
Therefore,
\begin{equation}
%\begin{split}
J (\rho,z) = 
\frac{1}{\sqrt{2}}
%\begin{pmatrix}\rho^{\tilde{p}} e^{\alpha iz} & 0 \\ 0 & \rho^{-\tilde{p}} e^{-\alpha iz} \end{pmatrix} \\ & \times \left(
\begin{pmatrix}
\rho^{2\tilde{p}} e^{i\alpha z}& e^{i\alpha z}g_0 +
\mathcal{O}\left(\rho^{2\tilde{p}+1}
 \right) \\
-e^{-i\alpha z}\hat{g}_0  +
\mathcal{O}\left(\rho^{2\tilde{p}+1}
 \right)   &  \rho^{-2\tilde{p}}e^{-i\alpha z}(2-\hat{g}_0 g_0 ) +
\mathcal{O}\left(\rho \right)
\end{pmatrix} 
 %  \right) \\ & \times \begin{pmatrix}\rho^{\tilde{p}} & 0 \\ 0 & \rho^{-\tilde{p}} \end{pmatrix} \, .\end{split}
\end{equation}
As $g_0$ and $\hat{g}_0$ are proportional to $\rho^{2\tilde{p}}$, it is clear that the entries of $J$, $J^{-1}$, and $\parrh J$ are all meromorphic when $\rho \rightarrow 0$.

We can now proceed to the second step. Recall that in a particular gauge the Lax pair can be written as in eq.~(\ref{laxpair}),
\begin{align}
L_1 &= \partial_{\rho} + i \lambda \parz + \frac{1}{\rho}\lambda \partial_\lambda + i \lambda J^{-1} \parz J \, , \\
L_2 &= \parz -i \lambda \partial_{\rho} + \frac{i}{\rho} \lambda^2 \partial_{\lambda} -i \lambda J^{-1}\partial_{\rho} J \, .
\end{align}
In order to get to eq.~(\ref{PIIIsystem}) we first need to perform a gauge transformation and go to a $z$-independent frame. From eq.~(\ref{SL2Solution}) it is clear that
\begin{equation}
J^{-1} \parz J = J^{-1} \begin{pmatrix}
-\tilde a i & 0 \\ 0 & \tilde a i
\end{pmatrix} J+  \begin{pmatrix}
a i& 0 \\ 0 & -ai
\end{pmatrix} \, .
\end{equation}
Therefore, 
\begin{equation}
\begin{pmatrix}
e^{{a } iz} & 0 \\ 0 & e^{-{a }i z}
\end{pmatrix}
L_1 \begin{pmatrix}
e^{-{a }i z} & 0 \\ 0 & e^{{a } iz}
\end{pmatrix} = \parrh  + \frac{1}{\rho} \lambda \partial_\lambda + i \lambda J^{-1} \begin{pmatrix}
-\tilde{a}i& 0 \\ 0 & \tilde{a}i
\end{pmatrix} J |_{z=0} \, .
\end{equation}
The same gauge transformation gives
\begin{align}
\begin{pmatrix}
 ie^{{a } iz} & 0 \\ 0 & e^{-{a } iz}
\end{pmatrix}
L_2 \begin{pmatrix}
e^{{-a }i z} & 0 \\ 0 & e^{{a } iz}
\end{pmatrix}= 
\lambda \parrh - \frac{\lambda^2}{\rho} \partial_\lambda + \lambda J^{-1} \parrh J |_{z=0} +  \begin{pmatrix}
{a } & 0 \\ 0 & -{a }
\end{pmatrix} \, .
\end{align}
It follows that 
\begin{equation}
\begin{split}
B &=  J^{-1} 
\begin{pmatrix}
-\tilde{a} & \\ 0 & \tilde{a}
\end{pmatrix} J |_{z=0}\\
C &=  \begin{pmatrix}
-{a } & 0 \\
0 & {a } 
\end{pmatrix}  \\
D &= - J^{-1} \parrh J|_{z=0} \, .
\end{split} 
\end{equation}
In particular, using the knowledge gained in the first step, we can conclude that the entries $D_{12}$, $D_{21}$, $B_{12}$, $B_{21}$ are meromorphic in $\rho$ as $\rho \rightarrow 0$. In more detail, from eq.~(\ref{transcendent}), provided $\tilde{p}>0$, we have
\begin{equation}
D_{12}= -4\tilde{p} \frac{\tilde{a}}{2^{2\tilde{p}}\tilde{p}!} \frac{1}{\rho} + \mathcal{O} (\rho^0) \, \quad B_{12}=-2 \frac{\tilde{a}^2}{2^{2\tilde{p}}\tilde{p}!} + \mathcal{O}(\rho)  \, .
\end{equation}
\begin{equation}
f= -\frac{1}{2} \frac{D_{12}}{B_{12}} = -\frac{\tilde{p}}{\tilde{a}} \frac{1}{\rho}  +\mathcal{O}(\rho^0)  = \frac{1}{2} \frac{n}{l^2} \frac{1}{\rho}  +\mathcal{O}(\rho^0)  = -\frac{\alpha}{\gamma}\frac{1}{\rho}  +\mathcal{O}(\rho^0) \, ,
\end{equation}
in agreement with the result of the Frobenius analysis eq.~(\ref{a_0}). If $\tilde{p}=0$ we find instead
$$
f=0 + \mathcal{O}(\rho^{-1}) \, ,
$$
which is also consistent with the Frobenius analysis.

\subsection{Characterization of monodromy data}
We now prove Theorem~\ref{MonodromyTheorem} that states that the solutions parametrised in Theorem~\ref{MainTheorem} are the solutions whose monodromy data $\mathcal{M}$ has trivial Stokes matrices and half-integral exponents of formal monodromy. First, assume that we are given the Riemann-Hilbert problem of  Theorem~\ref{MainTheorem}. We restrict to the case $D_6$ as $Q$ can be treated in a similar way. The patching matrix eq.~(\ref{MeromorphicPIII}) can be written as
\begin{equation}
P = 
\begin{pmatrix}
(\rho \lambda)^{\tilde {p }} & 0 \\
0 & (\rho \lambda)^{-\tilde {p }} 
\end{pmatrix}
\begin{pmatrix}
e^{au}&0 \\
0& e^{-au}
\end{pmatrix}
\begin{pmatrix}
 c_0 & \tilde c_1 \\
  c_1 & \tilde c_0 
\end{pmatrix}
\begin{pmatrix}
e^{-\tilde au}& 0\\
0 & e^{+ \tilde au}
\end{pmatrix}
\begin{pmatrix}
(\frac{\lambda}{\rho})^{{p}} & 0 \\
0 & (\frac{\lambda}{\rho})^{-{p}} 
\end{pmatrix},
\end{equation}
This patching matrix is meant to relate two frames $\hat F^{0}$ and $\hat  F^{\infty}$ holomorphic near $\lambda = 0$ and $\lambda = \infty$ respectively. In order to eliminate the diagonal matrices,  redefine the frames so that
\begin{equation}
\hat F^{(0)} \mapsto \hat F^{(0)}
\begin{pmatrix}
 e^{\tilde au}(\frac{\lambda}{\rho})^{{-p}} & 0  \\
 0 & e^{-\tilde au} (\frac{\lambda}{\rho})^{{p}}
\end{pmatrix}
\end{equation}
near $\lambda = 0$, and similarly for the frame near $\lambda = \infty$. Near $\lambda = 0$ we have
\begin{equation}
\hat F^{(0)}
\begin{pmatrix}
 e^{\tilde au}(\frac{\lambda}{\rho})^{{-p}} & 0  \\
0 &  e^{-\tilde au} (\frac{\lambda}{\rho})^{{p}}
\end{pmatrix}
\sim
\hat Y^{(0)} \exp \left( \tilde a \frac{\rho\lambda}{2} \begin{pmatrix}
1 & 0 \\ 0  & -1
\end{pmatrix} -p\log  \lambda   \begin{pmatrix}
1 & 0 \\ 0  & -1
\end{pmatrix} \right) \, ,
\end{equation}
where we reabsorbed $\lambda$-independent factors of $\hat F^{(0)}$ in $\hat Y^{(0)}$. This is of the form of eq.~(\ref{formalsolutions}), and since the asymptotic expansion for solutions of the linear system eq.~(\ref{linearsystem}) is unique it must be the same as eq.~(\ref{linearsystem}). By construction,  $\hat Y^{(0)}$ is holomorphic near $\lambda=0$ and therefore the Stokes matrices are trivial. We conclude that 
\begin{equation}
C:=\begin{pmatrix}
 c_0 & \tilde c_1 \\
  c_1 & \tilde c_0 
\end{pmatrix} 
\end{equation}
is by definition the connection matrix. We also see directly that the exponents formal monodromy are $\pm p$ which are half-integral (or of course integral) by our earlier discusion.  Our constructions  therefore maps to monodromy data $\mathcal{M}$ with trivial Stokes matrices and half-integer exponents of formal monodromy.

\section{The  Ward ansatz}
\label{sec:WardAnsatz}

The \emph{Ward ansatz} \cite{Ward:1982bf} constructs non-trivial examples of solutions by taking the  data to be upper triangular.  One can then solve the Riemann-Hilbert problem explicitly, or at least reduce the procedure to solving linear equations.  What is remarkable is that the solutions $J$ that are obtained cannot be reduced to being  upper triangular when the diagonal entries have nontrivial winding number.  Reducing to the $SL(2,\C)$ case,  we take  patching matrices $P$ of the  form 
\begin{equation}
P= \begin{pmatrix}
\rho^{s} \lambda^r e^{\sigma (u)} & \rho^{r} \lambda^{s}   \gamma(u) \\
0 & \rho^{-s} \lambda^{-r} e^{-\sigma (u)}
\end{pmatrix} .
\label{WardAnsatz}
\end{equation}
Here $\sigma$ and $\gamma$ are holomorphic functions of $u$, and with respect to eq.~(\ref{patching}) we have set $r=p+\tilde p $, $s=p - \tilde p$. We must  assume $r\geq 0$ in order that there is not a line subbundle of positive degree (which would contradict triviality of the bundle on a line).

The original work is \cite{Corrigan:1978}. Details closer to our approach can be found for example in \cite{Ward:1990}, p.398 et seq. The computation of the $J$-matrix based on the procedure outlined therein leads to the following theorem, which generalizes proposition A.1 of \cite{Masuda:2007}.
\begin{thm}  Let $A=A(\rho,z)$ and $A_\rho=\p_\rho A$, $A_{z}=\p_z A$   satisfying
\begin{equation}
\left( \frac{1}{\rho} + \parrh \right) A_\rho  + \parz A_z = 0,
\label{SDMaxwell}
\end{equation}
and  let $\Delta_k=\Delta_k(\rho,z)$  solve
\begin{equation}
\begin{split}
\left( \parrh  - \frac{k}{\rho} \right) \Delta_k = -i (\parz + A_z) \Delta_{k+1}, \\ \left( \parrh + A_\rho + \frac{k+1}{\rho} \right) \Delta_{k+1} = -i\parz \Delta_k.
\end{split}
\label{propass1}
\end{equation}
Define
\begin{equation}
\tau_r^s \equiv \det (t_r^s), \quad  t_r^s\equiv \begin{pmatrix}
\Delta_{s-r+1}  & \Delta_{s-r+2} & \cdots & \Delta_s \\
\Delta_{s-r+2} & \Delta_{s-r+3} & \cdots & \Delta_{s+1} \\
\vdots & \vdots & \ddots & \vdots \\
\Delta_{s} & \Delta_{s+1} & \cdots & \Delta_{s+r-1}
\end{pmatrix}.
\label{masudadeterminants}
\end{equation}
Then, provided $\tau_r^s \neq 0$,
\begin{equation}
J (\rho , z)= \frac{1}{\tau_r^s} 
\begin{pmatrix}
\rho^{s}\tau_{r}^{s+1} & \rho^{r}\tau_{r+1}^{s} \\
\rho^{-r} \tau_{r-1}^s & \rho^{-s} \tau_{r}^{s-1}
\end{pmatrix}
\label{Jmatrix}
\end{equation}
is a solution of the stationary axisymmetric SDYM equation, eq.~(\ref{SDYM-J}).
\label{MasudaProp}
\end{thm}

$A$ is an abelian analogue of $\log J$, (so that $A_\rho$, $A_z$ are abelian counterparts of $J^{-1} \parrh J$ and $ J^{-1} \parz J$ in eq.~(\ref{SDYM-J})). It arises from viewing $e^{\sigma (u) }$ as a $GL (1, \C)$-patching matrix and by following the steps in the proof of Theorem~1. In the abelian case, eq.~(\ref{Jfactorization}) amounts to a splitting
\begin{equation}
\sigma (\rho/2 (\lambda + 1/ \lambda)+ iz) =\sigma_{\infty} (\lambda , \rho, z) - \sigma_0 (\lambda, \rho, z)
\label{sigmasplitting}
\end{equation}
where $\sigma_0 (\lambda)$ is holomorphic in a neighborhood of $\lambda = 0$, whereas $\sigma_\infty (\lambda)$ is holomorphic in a neighbourhood of $\lambda = \infty$. As $V_1$, $V_2$ annihilate $\sigma$ restricted to a line, it follows that $V_1 \sigma_0 = V_1 \sigma_{\infty}$, $V_2 \sigma_0 = V_2 \sigma_{\infty}$, and by the Liouville-type argument that both expressions are at most linear in $\lambda$. Using the freedom in eq.~(\ref{sigmasplitting}) we can remove the constant terms. Thus
\begin{equation}
V_1 \sigma_0 = i A_z \lambda, \quad V_2 \sigma_0 = -i A_\rho \lambda.
\end{equation}
and so 
\begin{equation}
(V_1 + i A_z \lambda ) e^{-\sigma_0} = (V_2 - i A_\rho \lambda ) e^{-\sigma_0} = 0 .
\end{equation}
Consequently
\begin{equation}
[V_1 + i A_z \lambda, V_2 - i A_\rho \lambda ] = 0,
\end{equation}
which is equivalent to eq.~(\ref{SDMaxwell}). 

The $\Delta_k$s arise from the following Laurent expansion in $\lambda$ in an annulus surrounding $|\lambda |=1$
\begin{equation}
\exp (-\sigma_{\infty}-\sigma_0) \gamma (\lambda , \rho, z) = \sum_{i \in \mathbb{Z}} \Delta_{-i}(\rho,z) \lambda^{i}.
\label{laurentgamma}
\end{equation}
Eq.~(\ref{propass1}) is then a direct consequence of the fact that the vector fields in eq.~(\ref{laxvector}) annihilate $\gamma (u)$. 

More geometrically, define the line bundles  $L\rightarrow \T(U)$ by its transition function $ e^{\sigma (u)}$, and $\CO(r)\rightarrow\CP^1$ by $\lambda^{-n}$ so that it has Chern class $n$. Then the patching matrix in eq.~(\ref{WardAnsatz}) represents the bundle $E$  as an extension of $L(-r):=L\otimes \CO(-r)$ by its dual.  Thus   $E$ fits into the following short exact sequence on $\mathbb{T}(U)$
\begin{equation}
0 \rightarrow L(-r) \rightarrow E \rightarrow L^{-1}(r) \rightarrow 0 .
\end{equation}
The Penrose-Ward transform identifies the line bundle $L$ with the stationary-axisymmetric self-dual Maxwell field fields with components $A_z$ and $A_\rho$ on the reduced space-time coordinatized by $(\rho, z)$. On the other hand, the off diagonal entry in $\lambda^m \gamma (u)$ can be seen as an element of $H^{1} (\mathbb{T}(U), L^2(-2r))$. The Penrose transform realizes such cohomology classes as  massless field of helicity $r-1$ coupled to the Maxwell field. Such a field has $2n+1$ components and these are the coefficients $\Delta_k$ for $|k| \leq r-1$, and the charged massless field equations in this stationary axisymmetric context are \eqref{propass1}. 

\paragraph{Painlev\'e III example.} As an example, we study the case in which $c_0=\tilde c_0=1$, $c_1=0$ in eq.~(\ref{SL2Solution}). Defining $r = {p }+\tilde{p}$, $s = -{p }+\tilde{p}$, we obtain
\begin{equation}
\begin{pmatrix}
\rho^s \lambda^r e^{ ({a } - \tilde{a}) u }  &  \rho^r\lambda^s e^{ (-{a } - \tilde{a}) u } \tilde c_1 \\
0& \rho^{-s} \lambda^{-r} e^{ (-{a } + \tilde{a}) u } 
\end{pmatrix},
\end{equation}
where
\begin{equation}
 u =  \frac{\rho}{2}\left(\lambda + \frac{1}{\lambda} \right) + iz.
\end{equation}
Based on the previous discussion, we first need to split $({a }-\tilde{a})u$ as
\begin{equation}
\sigma_\infty = ({a }-\tilde{a}) \frac{\rho}{2\lambda}, \quad \sigma_0 = -({a }-\tilde{a} ) \left( \frac{\rho}{2}\lambda + iz \right).
\end{equation}
In order to perform the Laurent expansion eq.~(\ref{laurentgamma}), we recall the well-known identity
\begin{equation}
e^{\frac{\rho}{2}\left( \lambda + \frac{1}{\lambda} \right)} = \sum_{i=-\infty}^{\infty} I_{-i} (\rho) \lambda^i,
\label{Besselexpansion}
\end{equation}
where the $I_i(\rho)$s are modified Bessel functions of the first kind. We therefore rearrange the LHS of eq.~(\ref{laurentgamma}) as follows,
\begin{equation}
\tilde c_1 e^{-\sigma_\infty -\sigma_0}e^{(-\tilde{a}-{a }) u} = \tilde c_1 \mathrm{exp} \left( - \rho \sqrt{{a } \tilde{a}} \left( \tilde{\lambda} + \frac{1}{\tilde{\lambda}}\right)  \right) e^{-2\tilde{a} i z},
\end{equation}
where 
\begin{equation}
\tilde{\lambda} = \sqrt{\frac{\tilde{a}}{a }}\lambda.
\end{equation}
It then follows directly from eq.~(\ref{Besselexpansion}) that
\begin{equation}
\tilde c_1 e^{-\sigma_\infty -\sigma_0}e^{(-\tilde{a}-{a }) u}   = \sum_{j \in \mathbb{Z}} \tilde c_1 e^{-2\tilde{a} iz} I_{-j} (-2\rho \sqrt{{a }\tilde{a}}) \left(\frac{\tilde{a} }{{a }}\right)^{j/2}  \lambda^{j} , 
\end{equation}
and thus
\begin{equation}
\Delta_j =  \tilde c_1 e^{-2\tilde{a} iz} I_{j} \left( -2\rho \sqrt{{a }\tilde{a}}\right) \left(\frac{{a }}{\tilde{a}}\right)^{\frac{j}{2}}.
\end{equation}
In terms of the constants $m$, $n$, $k$, $l$ (picking the square-roots $l=\tilde{a}$, $-4k=a$)
\begin{equation}
\Delta_j =  \tilde c_1 e^{-2 l iz} I_{j} \left( -\rho \sqrt{-16 k l}\right) \left(-4\frac{{ k }}{l}\right)^{j/2}.
\end{equation}
and
\begin{equation}
s= \frac{1}{2}\left(\frac{m}{k}-\frac{n}{l}\right) , \ r = \frac{1}{2}\left(\frac{m}{k} + \frac{n}{l}\right) \, .
\end{equation}
In the context of the Painlev\'e equations, solutions involving special functions (Bessel functions in the case of $P_{III}$) are called \emph{classical transcendental solutions}. The classical transcendental solutions were classified in \cite{Masuda:2007} and given in terms of $J$-matrices of precisely the form of eq.~(\ref{Jmatrix}), see in particular theorem 4.2 of \cite{Masuda:2007}. These solutions coincide with ours up to a redefinition of the constants\footnote{The classical transcendental solutions presented in \cite{Masuda:2007} are essentially given by determinants as in (\ref{masudadeterminants}) with entries 
\begin{equation}
\phi_j = (-2 \eta_0)^{-j} e^{-\eta_0 z' + \eta_\infty \tilde{z}' } \tilde{w}^{\nu + 1- j} \rho^{-(\nu+1-j)} \psi_{\nu+1-j},
\end{equation}
where
\begin{equation}
\psi_{\nu} = 
\begin{cases}
c_1 J_{\nu} + c_2 Y_{\nu}, & 4 \eta_0 \eta_\infty = +1 \\
c_1 I_{\nu} + c_2 I_{-\nu} & 4 \eta_0 \eta_\infty = -1 \, ,
\end{cases}
\end{equation}
and
\begin{equation}
\begin{aligned}
\nu + 1 &= \frac{1}{2} \left( \frac{n}{l} + \frac{m}{k} \right) \\
\eta_\infty = 2l\,, & 
\qquad\eta_0 = -2k \,. \\ 
\end{aligned}
\end{equation}
The coordinate $\tilde{w}$ is related to $\rho$ by
$\tilde{w} = \rho e^{i \phi},
$ where $\phi$ is the the angular variable in the space-time cylindrical polar coordinates, and a combination of $z'$ and $\tilde z'$ gives our parameter $z$. The factor $\tilde{w}^{\nu + 1- j} \rho^{-(\nu+1-j)}$,  a power of $e^{i \phi}$, factors out from the $J$-matrix as well as exponentials containing $z$ or $z',\tilde{z}'$, and none  contributes to the transcendent.   With a rescaling of the transcendent we can set  $4 \eta_0 \eta_\infty =16 k l = \pm 1$  and it can then be checked that the solutions of \cite{Masuda:2007} meromorphic at $\rho=0$ agree with the ones we have obtained.}.

Our solutions reproduce all classical transcendental solutions meromorphic at $\rho=0$\footnote{This can also be checked from the point of view of the B\"acklund transformations of $P_{III}$.  This determines  the classical transcendental solutions as paricular loci in the parameter space of the constants $(\alpha, \beta, \gamma, \delta)$, see for example \cite{Masuda:2005}. The action of the B\"acklund transformations on the twistor data are given in \cite{Mason:1988}.}.

\section*{Acknowledgements} AF would like to thank the JP Stiftung, St. Gallen, for providing financial support during the completion of this work. LJM acknowledges support from EPSRC grant EP/M018911/1 and useful conversations with Peter Clarkson.

\bibliographystyle{JHEP}  
\bibliography{BibTex_Library}

\end{document}